\documentclass[aps,10pt,twocolumn,english,groupedaddress, superscriptaddress]{revtex4-1}
\usepackage[utf8]{inputenc}
\usepackage{gensymb}
\usepackage{textcomp}
\usepackage{amsmath,amsthm,mathtools}
\usepackage{amssymb}
\usepackage{graphicx}
\usepackage{babel}
\usepackage[normalem]{ulem}
\usepackage{url}
\usepackage{appendix}
\usepackage{algorithm}
\usepackage{algorithmic}
\usepackage{cancel}
\usepackage{pbox} 

\usepackage[colorinlistoftodos]{todonotes}
\definecolor{blue}{rgb}{0,0,1}
\newcommand{\revise}[1]{\textcolor{black}{#1}}

\begin{document}


\title{Superstatistical approach to air pollution statistics}

\author{Griffin Williams}
\affiliation{School of Mathematical Sciences, Queen Mary University of London, London E1 4NS, United Kingdom}

\author{Benjamin Sch\"afer}
\email{b.schaefer@qmul.ac.uk}
\affiliation{School of Mathematical Sciences, Queen Mary University of London, London E1 4NS, United Kingdom}

\author{Christian Beck}
\email{c.beck@qmul.ac.uk}
\affiliation{School of Mathematical Sciences, Queen Mary University of London, London E1 4NS, United Kingdom}

\begin{abstract}
Air pollution by Nitrogen Oxides (NOx) is a major concern in large cities as it has severe adverse health effects.  
However, the statistical properties of air pollutants are not fully understood.
Here, we use methods borrowed from non-equilibrium statistical mechanics to construct suitable superstatistical models for air pollution statistics. In particular, we
analyze time series of Nitritic Oxide ($NO$) and Nitrogen Dioxide ($NO_2$) concentrations recorded at several locations throughout Greater London. We find that the probability distributions of concentrations have heavy tails and that the dynamics is well-described
by $\chi^2$-superstatistics for $NO$ and inverse-$\chi^2$-superstatistics for $NO_2$. Our results can be used
to give precise risk estimates of high-pollution situations and pave the way to mitigation strategies.
\end{abstract}

\maketitle

Complex driven non-equilibrium systems with time scale separation
are often well-described by superstatistical methods, i.e. by
mixing several dynamics on distinct time scales and constructing  effective statistical mechanics models out of this mixing \cite{beck-cohen, BCS}. The intensive parameter $\beta$ that fluctuates in a superstatistical way can be an inverse temperature of the system, a fluctuating diffusion constant, or simply a local variance parameter in a given time series generated by the complex system under consideration. This formalism is in particular
relevant for heterogeneous spatio-temporally varying systems and has been successfully
applied to many areas of physics and beyond,
most notably Lagrangian turbulence \cite{beck2007},  defect turbulence \cite{bodenschatz}, wind velocity fluctuations \cite{rapisarda,weber2018wind}, share price dynamics \cite{jizba}, diffusion of complex bio-molecules \cite{metzler}, frequency fluctuations in power grids \cite{nature-energy}, rainfall statistics \cite{yalcin, demichel}, and many more.
Here, we apply a superstatistical analysis
to an important topic of high relevance, namely the spatio-temporally dynamics of air pollution in big cities. Our main example of pollutants considered in the following
are $NO$ and $NO_2$, but the method can be similarly applied to other substances.

$NO$ and $NO_{2}$ are examples of nitrogen oxides ($NO_{x}$), which are gaseous air pollutants that are primarily discharged through combustion \cite{Amster}.
In urban areas, the main cause of this is through automobile emissions \cite{Hamra}.
These chemicals have been shown to aggravate asthma and other respiratory symptoms and increase mortality. Further, short term exposure to $NO_{2}$ has been shown to correlate with ischaemic and hemorrhagic strokes \cite{Amster, Shah, USEPA}.
While $NO$ is a very important substance within Earth's ecosystem, for example regulating the development in plants and acting as a signaling hormone in certain processes in animals, it can also cause severe environmental damage in high concentrations \cite{Dom, Astier, Must}.
$NO$ may form acids, e.g. by chemically mixing with water ($H_{2}O$) to form Nitric Acid ($HNO_{3}$), or reacting further to $NO_{2}$ \cite{DHHS} it can cause acid rain \cite{USEPA}.
Though $NO_{x}$ are not thought to be carcinogenic, they indicate the presence of other harmful air pollutants \cite{Hamra}.

The data analyzed here was taken from the 
publicly available London Air Quality Network (LAQN) website
\cite{laqn}. The site provides readings of pollutant levels at different locations throughout Greater London at different time intervals, with the 15-minute interval data analyzed here.  
Inspecting the measured concentration time series for both $NO$ and $NO_{2}$ reveals that both pollutants' concentrations vary on multiple time scales, see Fig.~1.
\begin{figure*}
        \includegraphics[width=0.9\textwidth]{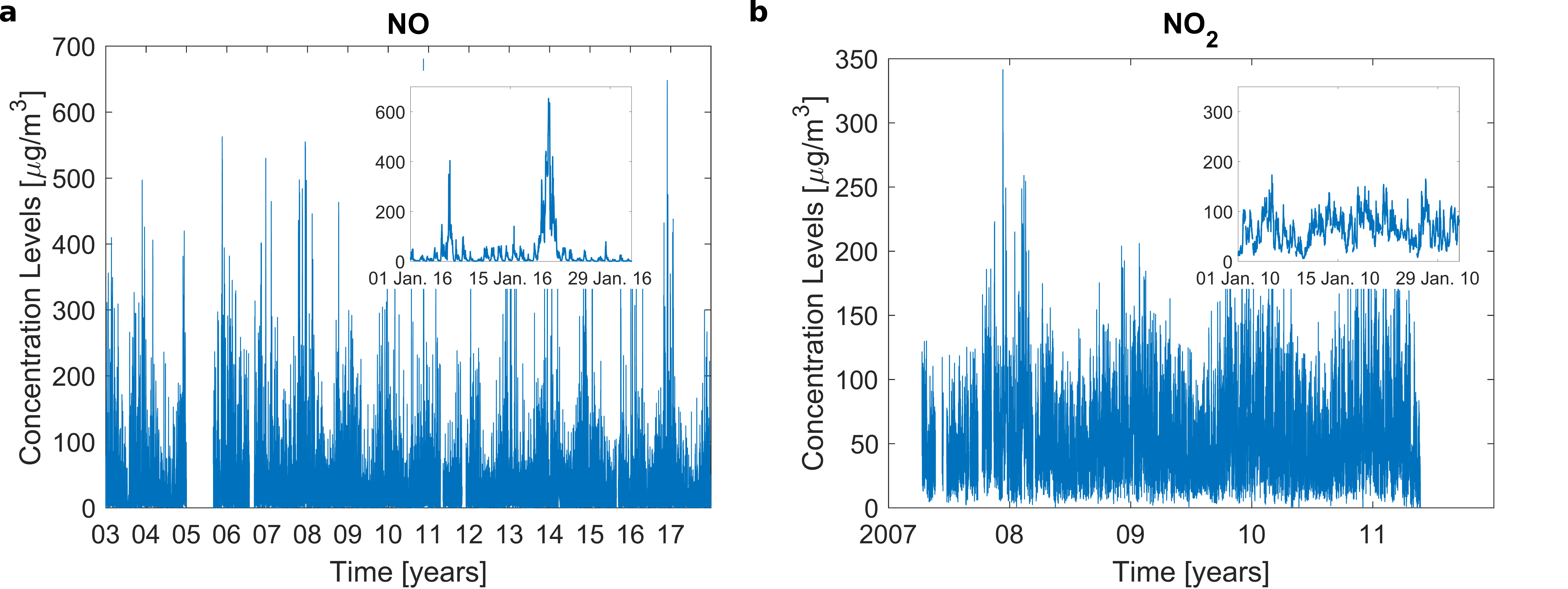}
        \caption{The concentration levels vary on several distinct time scales. a: We display the concentration of $NO$, recorded at the Sir John Cass School location (London), measured in 15-minute intervals. The inset reveals large variations within one month.
        b: We plot the concentration of $NO_2$, recorded at the North Street location (London), measured in 15-minute intervals. The inset again gives the concentrations for one month.
        For both pollutants we note seasonal changes, weekly changes and intermittent fluctuations.}
        \label{Fig1}
\end{figure*}

Concentrating on the case of $NO$ for now, we notice that the probability distribution of the measured concentration exhibits power-law tails, see Fig.~2. The distribution is
well-fitted by a $q$-exponential \cite{tsallis}
of the form
\begin{equation}
p(u)=(2-q)\lambda_{q} \left[1+(q-1)\lambda_{q}u\right]^{\frac{1}{1-q}},
\end{equation}
where $u$ are the concentration levels [$\mu g/m^3$] of the pollutant, and the $q$- and $\lambda_{q}$-values are parameters. $q$ can be
regarded as an entropic index \cite{tsallis, hanel2011, jizba2019} and it is related to $\lambda_{q}$ and the mean $\mu$ by
\begin{equation} \label{Eq2}
\lambda_{q}=\frac{1}{\mu(3-2q)},
\end{equation}
see Appendices A and B for more details.
\begin{figure}[h!]
        \includegraphics[width=0.45\textwidth]{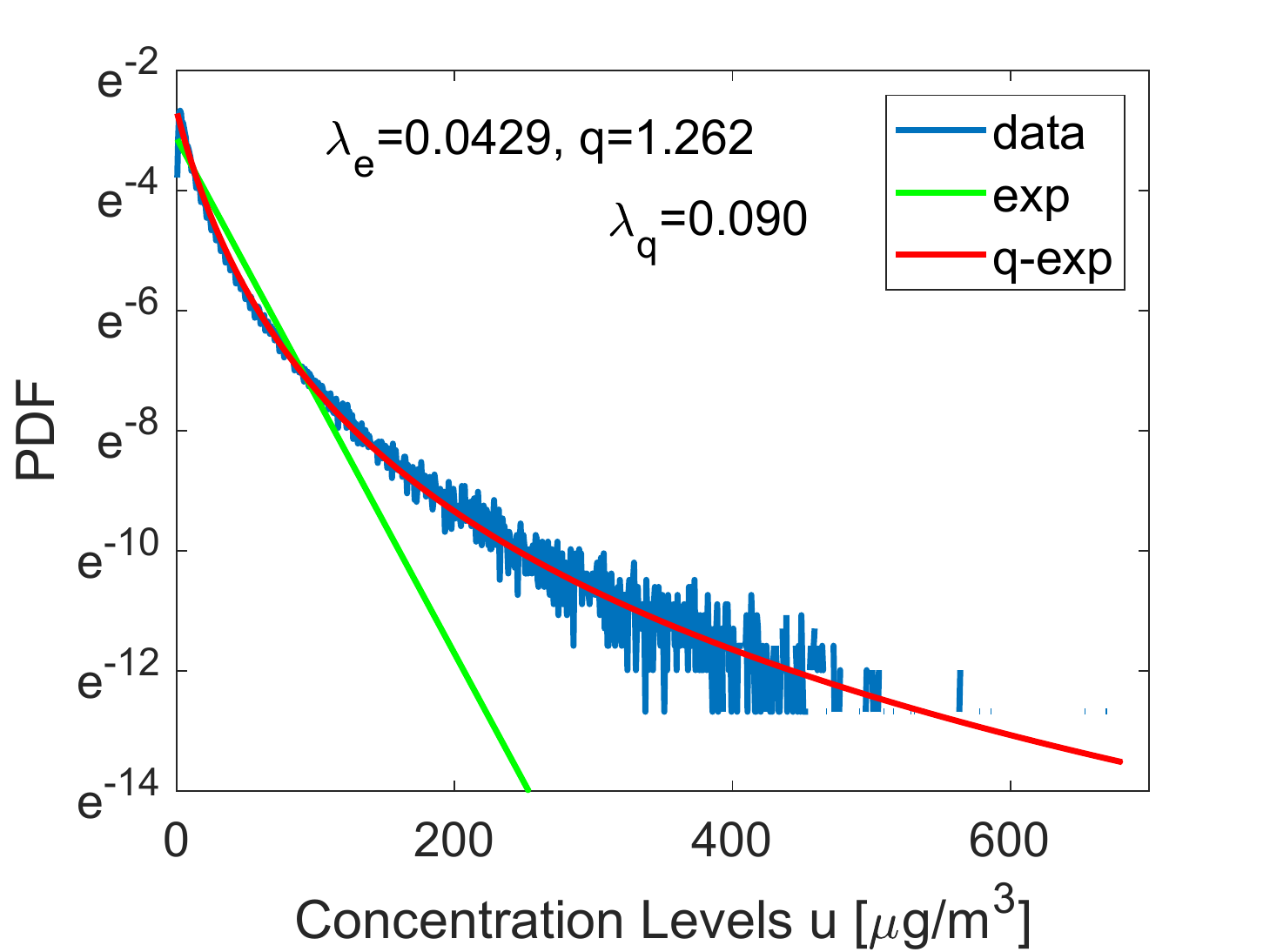}
        \caption{Pollution statistics is not exponentially distributed, but rather $q$-exponentially distributed.  We display the histogram of the $NO$ data, together with the best exponential fit (green) and a $q$-exponential fit (red). Also displayed are the value of the exponential parameter $\lambda_e$ and the $q$- and $\lambda_{q}$-parameters for the $q$-exponential distribution.  }
        \label{Fig2}
\end{figure}

The basic idea of the superstatistical modelling approach is that a given time series, generated by a complex driven non-equilibrium system, obeys a stochastic differential equation (SDE) where the parameters of the SDE change randomly as well, but on a much longer time scale \cite{beck2001}. In our case,
the concentration time series, whilst being $q$-exponentially distributed as a whole, appears to be
non-stationary, in the sense that it can be divided into shorter time slices, of length $T$, that each have locally an exponential density with different relaxation constant $\lambda_e$ each, see Fig.~3.
These changes in the pollutants statistics make sense due to the changing environment, e.g. because of changing weather conditions or traffic flow varying from week to week and also across seasons.
Observing simple (exponential) statistics locally, while noting heavy tails in the aggregated statistics clearly suggests a superstatistical description.
One of the simplest local models is based on an exponential probability distribution of the form
\begin{equation} \label{Eq3}
p(u|\beta)=\beta \cdot exp\left(-\beta u\right),
\end{equation}
where $\beta=\lambda_{e}$ is a local relaxation parameter that fluctuates
on the larger superstatistical time scale.
\begin{figure*}
        \includegraphics[width=0.9\textwidth]{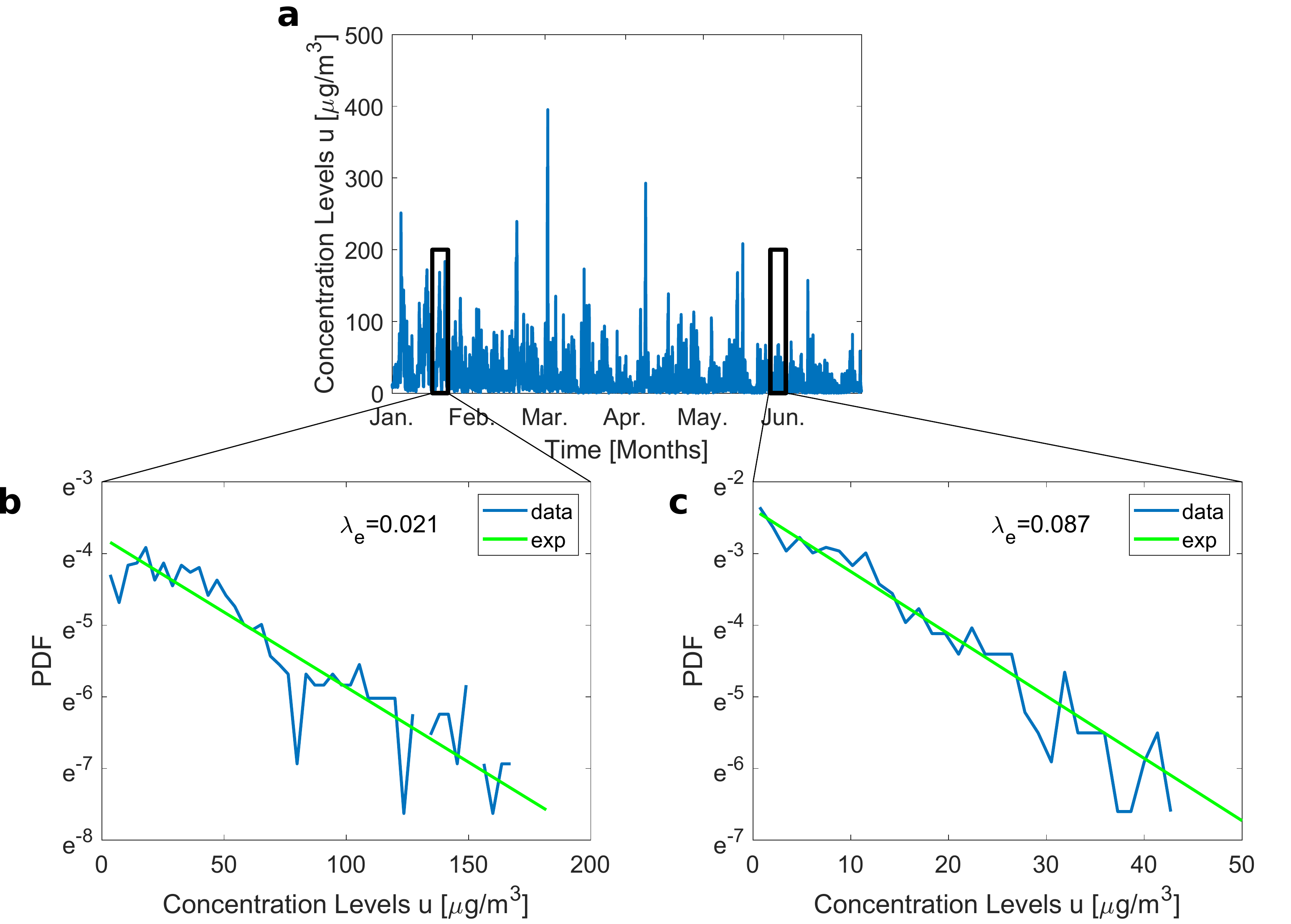}
        \caption{$NO$ concentrations locally follow exponential distributions. a: We plot the $NO$ concentration for several months in 2010, with two periods of length $\Delta t = T \approx 6~\text{days}$ highlighted, whose distributions are explored in panels b and c.  The histograms of panels b and c are fitted with exponential distributions and we note their respective $\lambda_{e}$-values. Note that the exponent in panel b is only one fourth of the exponent recovered in panel c, i.e. high pollution levels are much more likely to be observed in panel b, a time period in January, than in panel c, a time period in June. These different exponents quantify the observation that pollutant levels tend to be higher during winter than during summer, see also Fig.~\ref{Fig1}.} 
        \label{Fig3}
\end{figure*}

We determine this large time scale by computing the average local kurtosis $\kappa$ of a cell of length $\Delta t$, similarly as done in \cite{BCS}
for locally Gaussian distributions. We use a local average kurtosis defined as
\begin{equation} \label{Eq4}
\kappa\left(\Delta t\right)=\frac{1}{t_{max} - \Delta t}\int_{0}^{t_{max}-\Delta t} dt_{0}\frac{\langle\left(u-\bar{u}\right)^{4}\rangle_{t_{0},\Delta t}}{\langle\left(u-\bar{u}\right)^{2}\rangle_{t_{0},\Delta t}^{2}},
\end{equation}
where  $t_{max}$ is the full length of the time series. The notation $\langle\dots\rangle_{t_0, \Delta t}$ indicates the expectation for the time slice of length $\Delta t$ starting at $t_0$.  
Assuming local exponential distributions with kurtosis 9, we apply Equation (\ref{Eq4}) to determine the long time scale $T$ as the special time length  $\Delta t$ such that
\begin{equation}
\kappa\left(\Delta t\right)=\kappa\left(T\right)=9.
\end{equation}
We show an example of how $T$ is calculated in Appendix B, with a plot illustrating how the average local kurtosis depends on $\Delta t$.
Once the time slice length $T$ is calculated, we define an intensive parameter $\beta$ for each local cell,
by setting $\beta= \lambda_e =1/\langle u \rangle_{t_0,T}$. The distribution of this parameter, $f(\beta)$, is then obtained from a histogram, as shown in Fig.~4a. In
good approximation, we find $\beta$ to be
$\chi^{2}$ distributed, i.e.,
\begin{equation}\label{Eq7}
f(\beta)=\frac{1}{\Gamma\left(\frac{n}{2}\right)}\left(\frac{n}{2\beta_{0}}\right)^{\frac{n}{2}}\beta^{\frac{n}{2}-1}exp\left(-\frac{n\beta}{2\beta_{0}}\right),
\end{equation}
where $n$ is the number degrees of freedom and $\beta_{0}$ is the mean of $\beta$.

Integrating out the $\beta$-parameter, the marginal distribution $p(u)$ is calculated as
\begin{equation}
p(u)=\int_{0}^{\infty} p(u|\beta)f(\beta) \ d\beta, \label{Eq8}
\end{equation}
which evaluates to
\begin{equation}
p(u)=(2-q) \lambda_{q}\left[1+(q-1)\lambda_{q}u\right]^{\frac{1}{1-q}}, \label{Eq9}
\end{equation}
with
\begin{equation} \label{Eq10}
-\frac{n}{2}-1 = \frac{1}{1-q},
\end{equation}
and
\begin{equation}\label{Eq11}
\frac{1}{2}(q-1)\lambda_{q}=\frac{\beta_{0}}{n}.
\end{equation}
Eq. (\ref{Eq9}) is the $q$-exponential distribution, which we derived
from a superposition of local exponential distributions.

Fitting the $n$-parameter to find $f(\beta)$, as illustrated in Fig.~4a, we determine the $q$- and $\lambda_{q}$-parameters from Eqs. (\ref{Eq10}) and (\ref{Eq11}), respectively.  
Inserting the $\chi^2$-distribution $f(\beta)$, we use \eqref{Eq8} and \eqref{Eq9} to compute a probability density, which both approximates the fitted $q$-exponential and the data very well, successfully serving as a consistency check of the superstatistical approach, see Fig.~4b.

\begin{figure*}
    \includegraphics[width=.8\textwidth]{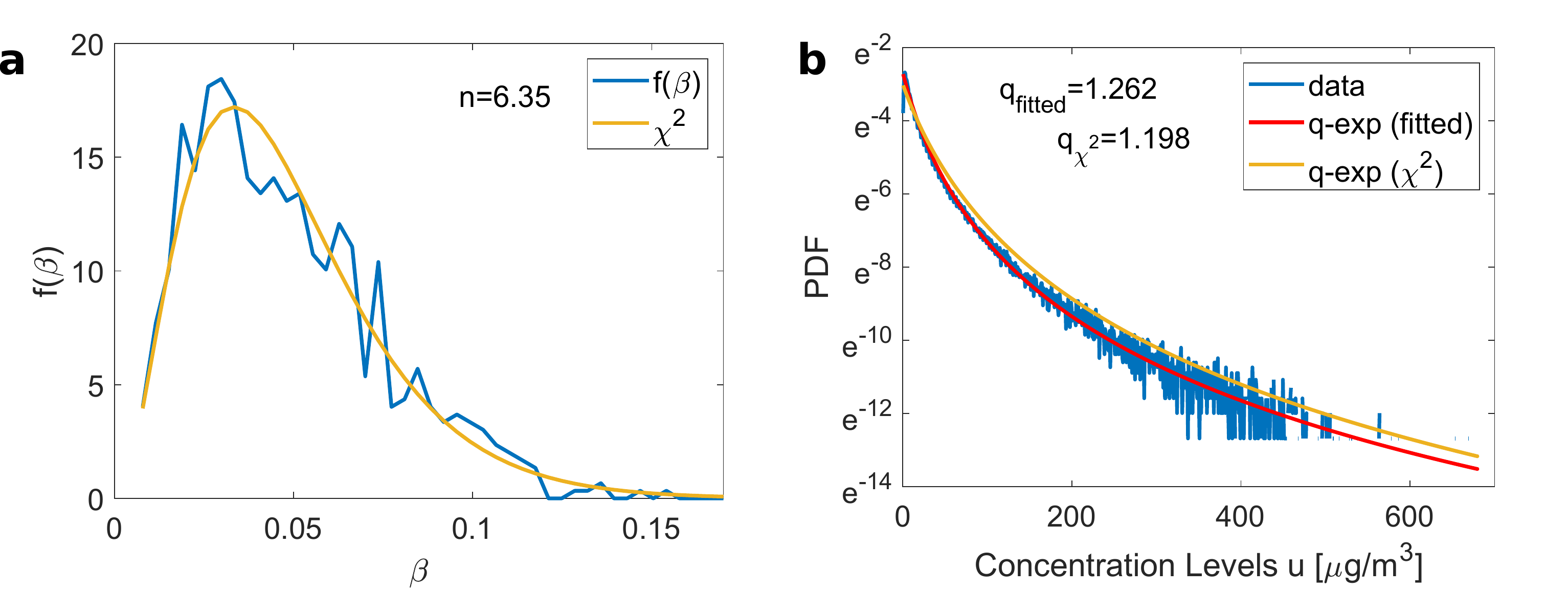}
    \caption{The superstatistical approach consistently describes the data. a: The distribution of local exponents of the  $NO$ data, here named $\beta$, follows a  $\chi^{2}$ distribution, one of three commonly observed universality classes in general superstatistical complex systems \cite{BCS}. b: Histogram of the $NO$ data, together with a $q$-exponential distribution with fitted $q$- and $\lambda_{q}$-values (red), and a $q$-exponential distribution with $q$- and $\lambda_{q}$-values derived from the parameters of the fitted $\chi^2$-distributed $f(\beta)$ (orange).}
        \label{Fig5}
\end{figure*}

\begin{figure*}
        \includegraphics[width=0.8\textwidth]{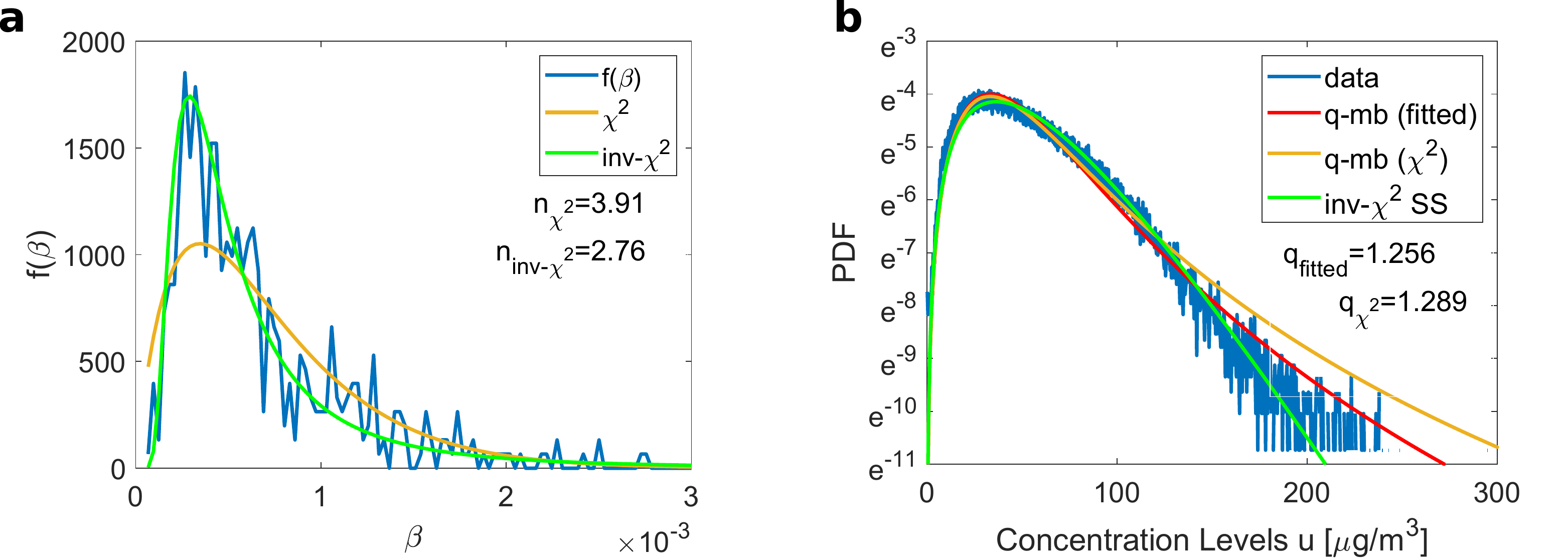}
        \caption{$NO_2$ concentrations are well described by inverse-$\chi^2$-superstatistics. a: The extracted
        distribution of the scale parameter $\beta$ for the $NO_2$ time series is very well fitted by an inverse-$\chi^{2}$ distribution (green). b: Histogram of the $NO_{2}$ data, together with a $q$-Maxwell-Boltzmann distribution with fitted $q$- and $\sigma_{q}$-values (red) and a $q$-Maxwell-Boltzmann with $q$- and $\sigma_{q}$-values superstatistically calculated from a $\chi^{2}$ distributed $f(\beta)$ (orange), and Eq. (\ref{Eq20}), which is calculated from inverse-$\chi^{2}$ superstatistics (green).
        The inverse-$\chi^2$-superstatistics fits the $NO_2$ data best. 
        }
        \label{Fig9}
\end{figure*}

We already noticed different statistical behavior when inspecting the trajectory of the $NO_2$ data in Fig.~1, as compared to that of $NO$. Following a similar procedure as for the $NO$ data, we find that for $NO_2$ the aggregated distribution is approximated by a $q$-Maxwell Boltzmann distribution, of the form
\begin{equation}
p(u)=\frac{1}{Z}u^{2}\sigma_{q}^{3/2}\left[1+(q-1)\sigma_{q}u^{2}\right]^{\frac{1}{1-q}},
\end{equation}
where $Z$ is the normalization factor given by
\begin{equation}
Z=\frac{\sqrt{\pi}\Gamma\left(\frac{5-3q}{2(q-1)}\right)}{4(q-1)^{3/2}\Gamma\left(\frac{1}{q-1}\right)} 
\end{equation}
and where $q$ is related to the scale parameter $\sigma_{q}$ and the mean $\mu$ by
\begin{equation}
\sigma_{q}=\left(\frac{2\left(q-1\right)^{3/2}\Gamma\left(\frac{1}{q-1}\right)}{\mu\sqrt{\pi}\left(2-q\right)\left(3-2q\right)\Gamma\left(\frac{5-3q}{2\left(q-1\right)}\right)}\right)^{2},
\end{equation}
see Appendix C for details. 


Analogously to the $NO$-case, we explain the aggregated distribution as a superposition of simple local distributions, here chosen as ordinary Maxwell-Boltzmann distributions. Each local Maxwell-Boltzmann distribution is defined as
\begin{equation}\label{Eq16}
p(u)=\sqrt{\frac{16}{\pi}}u^{2}\sigma_{mb}^{3/2}exp\left(-\sigma_{mb} u^{2}\right),
\end{equation}
where $\sigma_{mb}=: \beta$ is a scale parameter.  

We determine the long time scale $T$ by varying $\Delta t$ so that we locally obtain the kurtosis of a Maxwell-Boltzmann distribution, which is given as 
\begin{equation}\label{Eq17}
    \kappa\left(T\right)=\frac{15\pi^{2}+16\pi-192}{\left(3\pi-8\right)^{2}}\approx3.1082,
\end{equation}
see Appendix C for the calculation.
\revise{If the $\beta$ values again follow a $\chi^2$-distribution, the superpositioned Maxwell-Boltzmann distributions would lead to an exact $q$-Maxwell-Boltzmann distribution, see Appendix C for a detailed calculation.}

The results of our superstatistical analysis for
the $NO_{2}$ data are displayed in Fig.~5. \revise{The $\chi^2$-distribution is only a rough fit and the histogram of $\beta$ is instead best fitted by
an inverse-$\chi^{2}$-distribution}
\begin{equation}
    f(\beta)=\frac{\beta_{0}}{\Gamma\left(\frac{n}{2}\right)}\left(\frac{n\beta_{0}}{2}\right)^{\frac{n}{2}}\beta^{-\frac{n}{2}-2}exp\left(\frac{-n\beta_{0}}{2\beta}\right).
\end{equation}
This leads to a different superstatistics, compared to the case of $NO$.
Integrating the conditional probability $p(u|\beta)$, as in \eqref{Eq8} but with local Maxwell-Boltzmann distributions  $p(u|\beta)$ and an inverse-$\chi^2$-distribution $f(\beta)$, gives the marginal distribution  as
\begin{equation}\label{Eq20}
    p(u)=\sqrt{\frac{64}{\pi}}\frac{\beta_{0}}{\Gamma\left(\frac{n}{2}\right)}\left(\frac{n\beta_{0}}{2}\right)^{\frac{n+1}{4}}u^{\frac{n+3}{2}}K_{\frac{1-n}{2}}\left(u\sqrt{2n\beta_{0}}\right),
\end{equation}
where $K_{v}(z)$ is the modified Bessel function of the second order.  This inverse-$\chi^2$-superstatistical model leads to a very good description of the data, see Fig.~5, \revise{and it approximates $q$-Maxwell-Boltzmann distributions for medium concentrations but it not exactly the same. In fact, the PDF of the observed statistics decays like $p(u)\sim exp(-const \sqrt{u})$ for large $u$ values, see also Appendix C}.

To conclude, we have illustrated that superstatistical methods, originally
introduced in non-equilibrium statistical mechanics and applied to fully developed turbulent flows and other complex systems, find new applications to model the statistics of air pollution. We
find excellent agreement of simple superstatistical models with the experimentally
measured pollution data. A main result is that different pollutants obey
different types of superstatistics (in our case
$\chi^2$ for $NO$ and inverse-$\chi^2$ for $NO_2$). In fact, different
types of local dynamics also occur (in our case locally exponential density for $NO$ and
locally Maxwell-Boltzmann for $NO_2$). Once the precise superstatistical model
for a given pollutant has been identified, precise risk estimates
of high pollution situations can be given, by integrating the
tails of the probability distribution above a given threshold. These estimates could help
to design tailor-made thresholds for suitable policies to tackle the pollution problem in cities. We discuss some explicit examples in the Appendix D.
\revise{While the precise local statistics (exponential or Maxwell-Boltzmann) does significantly influence the likelihood to observe very low concentrations, the probability to observe high concentrations is determined by heavy tails in both cases.}

The superstatistical approach presented here can easily be 
generalized and extended. For example, we may formulate an explicit dynamical process to match the trajectories in time (in form of a superstatistical stochastic differential equation), which could lead to synthetic dynamical models and the possibility to employ short-term predictions of the pollution concentration. \revise{Further research should be devoted to understand how and why certain local statistics, such as exponential or Maxwell-Boltzmann distributions arise, and to potentially link these statistics to the physical and chemical properties of the pollutants.}




\begin{acknowledgements}
We acknowledge support from EPSRC Grant No. EP/N013492/1.
This project has received funding from the European Union’s Horizon 2020 research and innovation programme under the Marie Sklodowska--Curie grant agreement No 840825.
\end{acknowledgements}

\appendix
\section{Superstatistical cases}

These appendices provide calculations of some equalities used in the main text, such as relationships between mean, variance and kurtosis for the relevant distributions. We also display how the long time scale is determined and explicitly discuss how the knowledge of concentration probability distributions could be used to set pollution thresholds and policies.

\revise{We start by providing an overview of the different cases of local distributions (exponential or Maxwell-Boltzmann) and the $\beta$-distributions of the individual scale parameters:\\
\begin{tabular}{|c|c|c|}
\hline 
local distr./$\beta$ distr. & $\chi^{2}$ & inv. $\chi^{2}$\tabularnewline
\hline 
\hline 
exponential & \pbox{20cm}{ $q$-exp, \\ empirical $NO$ } 
& \pbox{20cm}{ Not observed here, \\ but see \cite{chen2008superstatistical}}\tabularnewline\hline 
Maxwell-Boltzmann & $q$-MB & empirical $NO_{2}$\tabularnewline
\hline 
\end{tabular}
}

\section{$NO$}

\revise{In the case of $NO$ concentrations, we approximate local concentrations as exponential distributions 
\begin{equation} \label{Eq3}
p(u)=\lambda_{e} \cdot exp\left(-\lambda_{e} u\right),
\end{equation}
and the aggregated statistics as $q$-exponentials
\begin{equation}
p(u)=(2-q) \lambda_{q}\left[1+(q-1)\lambda_{q}u\right]^{\frac{1}{1-q}}, 
\end{equation}
derived from a $\chi^2$ distribution of $\beta=\lambda_{e}$:
\begin{equation}\label{Eq7}
f(\beta)=\frac{1}{\Gamma\left(\frac{n}{2}\right)}\left(\frac{n}{2\beta_{0}}\right)^{\frac{n}{2}}\beta^{\frac{n}{2}-1}exp\left(-\frac{n\beta}{2\beta_{0}}\right).
\end{equation}
}

\subsection{Calculation of $\lambda_{q}$}
\revise{Here, we show how to express the scale parameter of $q$-exponentials $\lambda_q$ as a function of the mean of the distribution $\mu=\langle u\rangle$ and the $q$- parameter.
We start by computing the mean:}
\begin{equation}
    \begin{split}
        \mu=&\langle u\rangle=(2-q)\lambda_{q}\int_{0}^{\infty}u\left[1+(q-1)\lambda_{q}u\right]^{\frac{1}{1-q}} \ du \\
        =&\frac{(2-q)\Gamma(2)\Gamma\left(\frac{3-2q}{q-1}\right)}{\lambda_{q}(q-1)^{2}\Gamma\left(\frac{1}{q-1}\right)}\\
        =&\frac{1}{\lambda_{q}(3-2q)}
    \end{split}
\end{equation}
which gives
\begin{equation}
    \lambda_{q}=\frac{1}{\mu(3-2q)}.
\end{equation}
\subsection{Calculation of mean $\mu$ of Exponential Distribution}
\revise{We briefly recall the relationship between the exponential scale parameter $\beta=\lambda_e$ and the mean $\mu$ as}
\begin{equation}
    \begin{split}
        \mu=&\langle u \rangle=\beta\int_{0}^{\infty}u \ exp\left(-\beta u\right) \ du \\
                =&\frac{1}{\beta}.
    \end{split}
\end{equation}
\subsection{Calculation of kurtosis $\kappa$ of Exponential Distribution}
\revise{
In the main text, we determined the time scale $T$ as the time scale where the local kurtosis of the local exponential distribution takes on the value  $\kappa=9$.
Here, we provide the corresponding calculation} 
\begin{equation}
    \begin{split}
        \kappa=&\frac{\langle(u-\mu)^{4}\rangle}{\langle(u-\mu)^{2}\rangle^{2}}=\frac{\int_{0}^{\infty}(u-\mu)^{4}\beta exp\left(-\beta u\right) \ du}{\left(\int_{0}^{\infty}(u-\mu)^{2}\beta exp\left(-\beta u\right) \ du\right)^{2}} \\
        =&\frac{\int_{0}^{\infty}(u^{4}-4u^{3}\mu+6u^{2}\mu^{2}-4u\mu^{3}+\mu^{4})\beta exp\left(-\beta u\right) \ du}{\left(\int_{0}^{\infty}(u^{2}-2u\mu+\mu^{2})\beta exp\left(-\beta u\right) \ du\right)^{2}}\\
        =&\frac{24-24\mu\beta+12\mu^{2}\beta^{2}-4\mu^{3}\beta^{3}+\mu^{4}\beta^{4}}{4-8\mu\beta+8\mu^{2}\beta^{2}-4\mu^{3}\beta^{3}+\mu^{4}\beta^{4}}\\
        =&9,
    \end{split}
\end{equation}
using $\mu=\frac{1}{\beta}$.
\subsection{Superstatistical calculation of $q$-Exponential Distribution}
\revise{We show how integrating several local exponential distributions, whose exponents $\beta$ follow a $\chi^2$-distribution, leads to an overall $q$-exponential distribution: Each local distribution is given as}
\begin{equation}
    p(u|\beta)=\beta exp(-\beta u).
\end{equation}
\revise{Integrating over all of these distributions can be expressed as}
\begin{equation}
    \begin{split}
        p(u)=&\int_{0}^{\infty}p(u|\beta)f(\beta) \ d\beta \\
        =&\frac{1}{\Gamma\left(\frac{n}{2}\right)}\left(\frac{n}{2\beta_{0}}\right)^{\frac{n}{2}}\int_{0}^{\infty}\beta^{\frac{n}{2}}exp\left(-\beta\left(\frac{n}{2\beta_{0}}+u\right)\right) \ d\beta,
    \end{split}
\end{equation}
which we evaluate to
\begin{equation}
    p(u)=(2-q) \lambda_{q}\left[1+(q-1)\lambda_{q}u\right]^{\frac{1}{1-q}},
\end{equation}
if we identify
\begin{equation}
    -\left(\frac{n+2}{2}\right)=\frac{1}{1-q},
\end{equation}
\begin{equation}
    \frac{1}{2}(q-1)\lambda_{q}=\frac{\beta_{0}}{n},
\end{equation}
where $\beta_{0}$ is the mean of $\beta$.
\subsection{Determining the long time scale $T$} 
We determine the long time scale $T$ from the time series using Equ.~(4) from the main text given as
\begin{equation} 
\kappa\left(\Delta t\right)=\frac{1}{t_{max} - \Delta t}\int_{0}^{t_{max}-\Delta t} dt_{0}\frac{\langle\left(u-\bar{u}\right)^{4}\rangle_{t_{0},\Delta t}}{\langle\left(u-\bar{u}\right)^{2}\rangle_{t_{0},\Delta t}^{2}}.
\end{equation} 
The long time scale $T$ is defined  as $T:=\Delta t$ such that $\kappa(\Delta t)=9$. Here, we compute the average local kurtosis $\kappa$ as a function of the time window $\Delta t$ and thereby determine $T$.
\begin{figure}[h!]
        \includegraphics[width=0.5\textwidth]{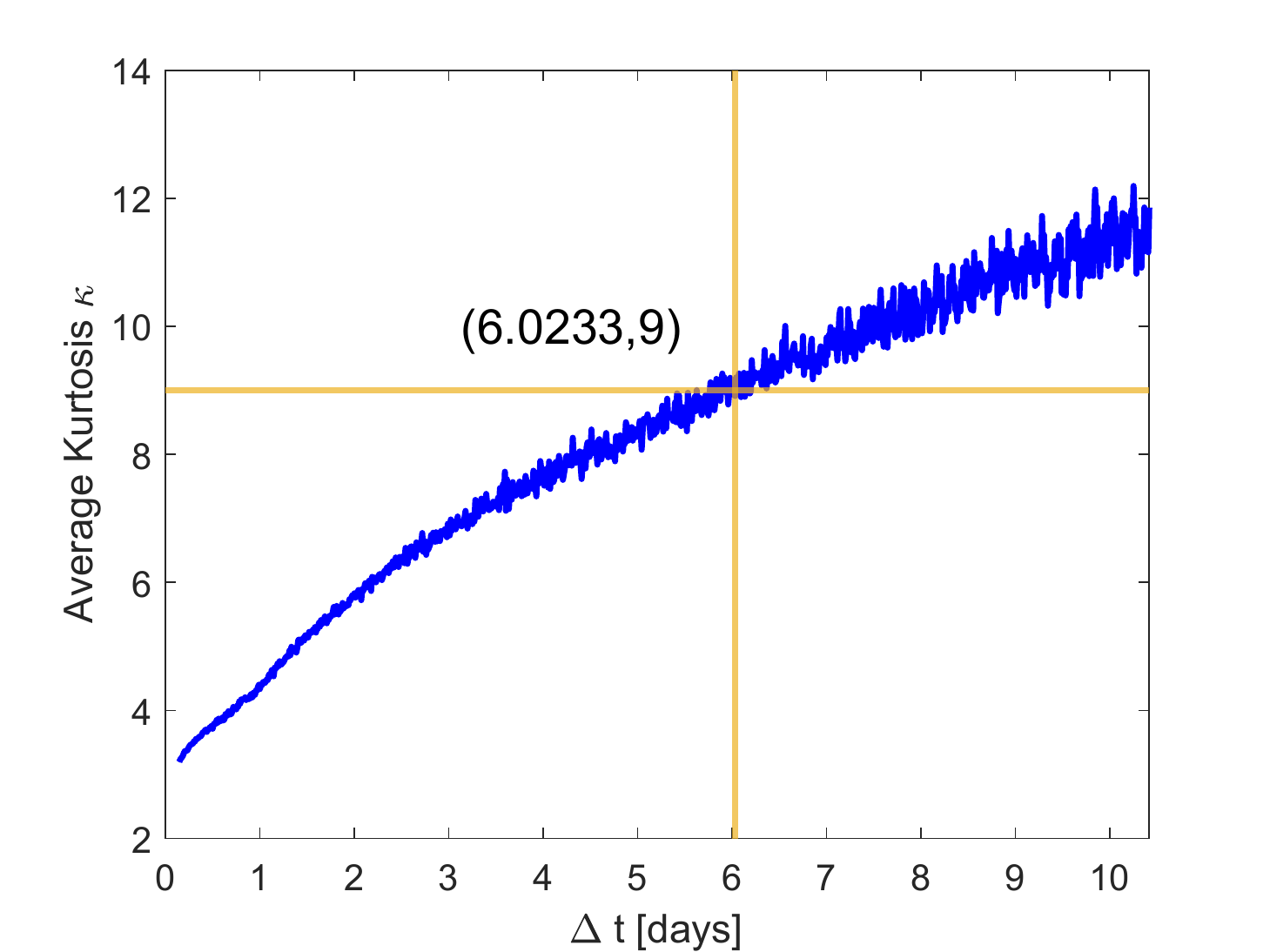}
        \caption{
        The average kurtosis $\kappa$ as given by Eq.(4) is plotted as a function of the time window $\Delta t$  (blue). 
        The crossing between the horizontal line at $\kappa=9$ (the kurtosis of an exponential distribution) and the $\kappa$ vs. $\Delta t$ curve gives the value for $\Delta t = T\approx 6$.}
        \label{Fig4}
\end{figure} 

\section{$NO_{2}$}

\revise{In the case of $NO_2$ concentrations, we approximate local concentrations as Maxwell-Boltzmann distributions 
\begin{equation}\label{Eq16}
p(u)=\sqrt{\frac{16}{\pi}}u^{2}\sigma_{mb}^{3/2}exp\left(-\sigma_{mb} u^{2}\right)
\end{equation}
and the aggregated statistics is a $q$-Maxwell-Boltzmann distribution
\begin{equation}
p(u)=\frac{1}{Z}u^{2}\sigma_{q}^{3/2}\left[1+(q-1)\sigma_{q}u^{2}\right]^{\frac{1}{1-q}}
\end{equation}
if $\sigma_{mb}$ is $\chi^2$ distributed.
These $q$-Maxwell-Boltzmann distributions would strictly arise if the scale parameters $\beta=\sigma_{mb}$ were following a $\chi^2$-distribution. Empirically, we observe that for our data, the $\beta$ values follow an inverse-$\chi^2$-distribution instead:
\begin{equation}
    f(\beta)=\frac{\beta_{0}}{\Gamma\left(\frac{n}{2}\right)}\left(\frac{n\beta_{0}}{2}\right)^{\frac{n}{2}}\beta^{-\frac{n}{2}-2}exp\left(\frac{-n\beta_{0}}{2\beta}\right).
\end{equation}
We obtain this inverse-$\chi^2$-distribution from a simple transformation of random variables when $\beta^{-1}$, rather than $\beta$, is $\chi^2$-distributed.
}

\subsection{Calculation of $\sigma_{q}$}
\revise{Here, we show how to express the scale parameter of $q$-Maxwell-Boltzmann distributions $\sigma_{q}$ as a function of the mean of the distribution $\mu=\langle u\rangle$ and the $q$-parameter.
We start by computing the mean:}
\begin{equation}
    \begin{split}
        \mu=&\langle u\rangle = \frac{\sigma_{q}^{3/2}}{Z}\int_{0}^{\infty}u^{3}\left[1+(q-1)\sigma_{q}u^{2}\right]^{\frac{1}{1-q}} \ du \\
        =&\frac{\Gamma(2)\Gamma\left(\frac{3-2q}{q-1}\right)}{2Z\sigma_{q}^{1/2}(q-1)^{2}\Gamma\left(\frac{1}{q-1}\right)}\\
        =&\frac{1}{2Z(2-q)(3-2q)\sigma_{q}^{1/2}}\\
        =&\frac{2(q-1)^{3/2}\Gamma\left(\frac{1}{q-1}\right)}{\sigma_{q}^{1/2}\sqrt{\pi}(2-q)(3-2q)\Gamma\left(\frac{5-3q}{q-1}\right)},
    \end{split}
\end{equation}
which gives
\begin{equation}
    \sigma_{q}=\left(\frac{2(q-1)^{3/2}\Gamma\left(\frac{1}{q-1}\right)}{\mu\sqrt{\pi}(2-q)(3-2q)\Gamma\left(\frac{5-3q}{q-1}\right)}\right)^{2}.
\end{equation}
\subsection{Calculation of mean $\mu$ of Maxwell-Boltzmann Distribution}
\revise{We briefly recall the relationship between the scale parameter $\beta=\sigma_{mb}$ of the Maxwell-Boltzmann distribution and the mean $\mu$ as}
\begin{equation}
    \begin{split}
        \mu=&\langle u \rangle= \sqrt{\frac{16}{\pi}}\beta^{3/2}\int_{0}^{\infty}u^{3}exp\left(-\beta u^{2}\right) \ du\\
        =& \sqrt{\frac{16}{\pi}}\beta^{3/2}\left(\frac{\Gamma(2)}{2\beta^{2}}\right)\\
        =&\frac{2}{\sqrt{\pi\beta}}
    \end{split}
\end{equation}
\subsection{Calculation of kurtosis $\kappa$ of Maxwell-Boltzmann Distribution}
\revise{In the main text, we determined the time scale $T$ as the time scale where the local kurtosis of the local Maxwell-Boltzmann distribution takes on the value  $\kappa\approx 3.1082$. Here, we provide the corresponding calculation} 
\begin{equation}
    \begin{split}
        \kappa=&\frac{\langle(u-\mu)^{4}\rangle}{\langle(u-\mu)^{2}\rangle^{2}}\\
        =&\frac{\sqrt{\frac{16}{\pi}}\beta^{3/2}\int_{0}^{\infty}\left(u-\mu\right)^{4}u^{2}exp(-\beta u^{2}) \ du}{\left(\sqrt{\frac{16}{\pi}}\beta^{3/2}\int_{0}^{\infty}\left(u-\mu\right)^{2}u^{2}exp(-\beta u^{2}) \ du\right)^{2}}\\
        =&\frac{\frac{\mu^{4}\sqrt{\pi}}{4\beta^{3/2}}-\frac{2\mu^{3}}{\beta^{2}}+\frac{9\mu^{2}\sqrt{\pi}}{4\beta^{5/2}}-\frac{4\mu}{\beta^{3}}+\frac{15\sqrt{\pi}}{16\beta^{7/2}}}{\sqrt{\frac{16}{\pi}}\beta^{3/2}\left(\frac{\mu^{2}\sqrt{\pi}}{4\beta^{3/2}}-\frac{\mu}{\beta^{2}}+\frac{3\sqrt{\pi}}{8\beta^{5/2}}\right)^{2}}\\
        =&\frac{15\pi^{2}+16\pi-192}{(3\pi-8)^{2}}\approx3.1082
    \end{split}
\end{equation}
\subsection{Superstatistical calculation of $q$-Maxwell-Boltzmann Distribution}
\revise{We show how integrating several local Maxwell-Boltzmann distributions, whose exponents $\beta$ follow a $\chi^2$-distribution, leads to an overall $q$-Maxwell-Boltzmann distribution: Each local distribution is given as}
\begin{equation}
    p\left(u|\beta\right)=\sqrt{\frac{16}{\pi}}u^{2}\sigma_{mb}^{3/2}exp\left(-\sigma_{mb}u^{2}\right). \label{here}
\end{equation}
\revise{Integrating over all of these distributions can be expressed as}
\begin{equation}
    \begin{split}
        p(u)=&\int_{0}^{\infty}p(u|\beta)f(\beta) \ d\beta \\
        =&\sqrt{\frac{16}{\pi}}\frac{1}{\Gamma\left(\frac{n}{2}\right)}\left(\frac{n}{2\beta_{0}}\right)^{\frac{n}{2}}u^{2} \\
        & \times  \int_{0}^{\infty}\beta^{\frac{n+1}{2}}exp\left(\beta\left(\frac{n}{2\beta_{0}}+u^{2}\right)\right) \ d\beta,
    \end{split}
\end{equation}
which can be evaluated to 
\begin{equation}
    p(u)\sim u^{2}\sigma_{q}^{3/2}\left[1+(q-1)\sigma_{q}u^{2}\right]^{\frac{1}{1-q}},
\end{equation}
if we identify 
\begin{equation}
    -\left(\frac{n+3}{2}\right)=\frac{1}{1-q},
\end{equation}
\begin{equation}
    \frac{1}{2}(q-1)\sigma_{q}=\frac{\beta_{0}}{n},
\end{equation}
where $\beta_{0}$ is the mean of $\beta$.
\subsection{Calculation of inverse-$\chi^{2}$ superstatistical distribution based on superimposed Maxwell-Boltzmann distributions}
\revise{Instead of following a $\chi^2$-distribution, our data indicates that the scale parameters $\beta$ of the local Maxwell-Boltzmann distribution rather follow an inverse-$\chi^2$-distribution. Here, we derive the probability density function for such a superposition.}
Taking the conditional density $p(u|\beta)$ as
given in (\ref{here}) and the inverse-$\chi^{2}$-distribution for the distribution   $f(\beta)$, the probability density $p(u)$ is given as
\begin{equation}
    \begin{split}
        p(u)=&\sqrt{\frac{16}{\pi}}\frac{\beta_{0}}{\Gamma\left(\frac{n}{2}\right)}\left(\frac{n\beta_{0}}{2}\right)^{\frac{n}{2}}u^{2} \\
        & \times \int_{0}^{\infty}\beta^{-\frac{(n+1)}{2}}exp\left(-\frac{n\beta_{0}}{2\beta}-\beta u^{2}\right) \ d\beta \\
        =&\sqrt{\frac{64}{\pi}}\frac{\beta_{0}}{\Gamma\left(\frac{n}{2}\right)}\left(\frac{n\beta_{0}}{2}\right)^{\frac{n+1}{4}}u^{\frac{n+3}{2}}K_{\frac{1-n}{2}}\left(u\sqrt{2n\beta_{0}}\right).
    \end{split}
\end{equation}
\revise{This PDF has still heavy tails, which decay as $p(u)\sim exp(-const \sqrt{u})$ for large $u$ values.}
\subsection{Extra Figures} 
We repeat Fig.~3 from the main text for local $NO_2$ concentrations, i.e. analyzing brief periods of the $NO_2$ trajectory.
Fig.~7 illustrates that in good approximation the local behavior follows Maxwell-Boltzmann distributions with fluctuating variance. Further, we also determine the long time scale $T$ for the local Maxwell-Boltzmann distributions by setting $T:=\Delta t$ for the local kurtosis $\kappa(\Delta t)\approx 3.1082$, following Eq.~(4) from the main text again.
See Figs.~7 and 8.
\begin{figure*}
\includegraphics[width=0.85\textwidth]{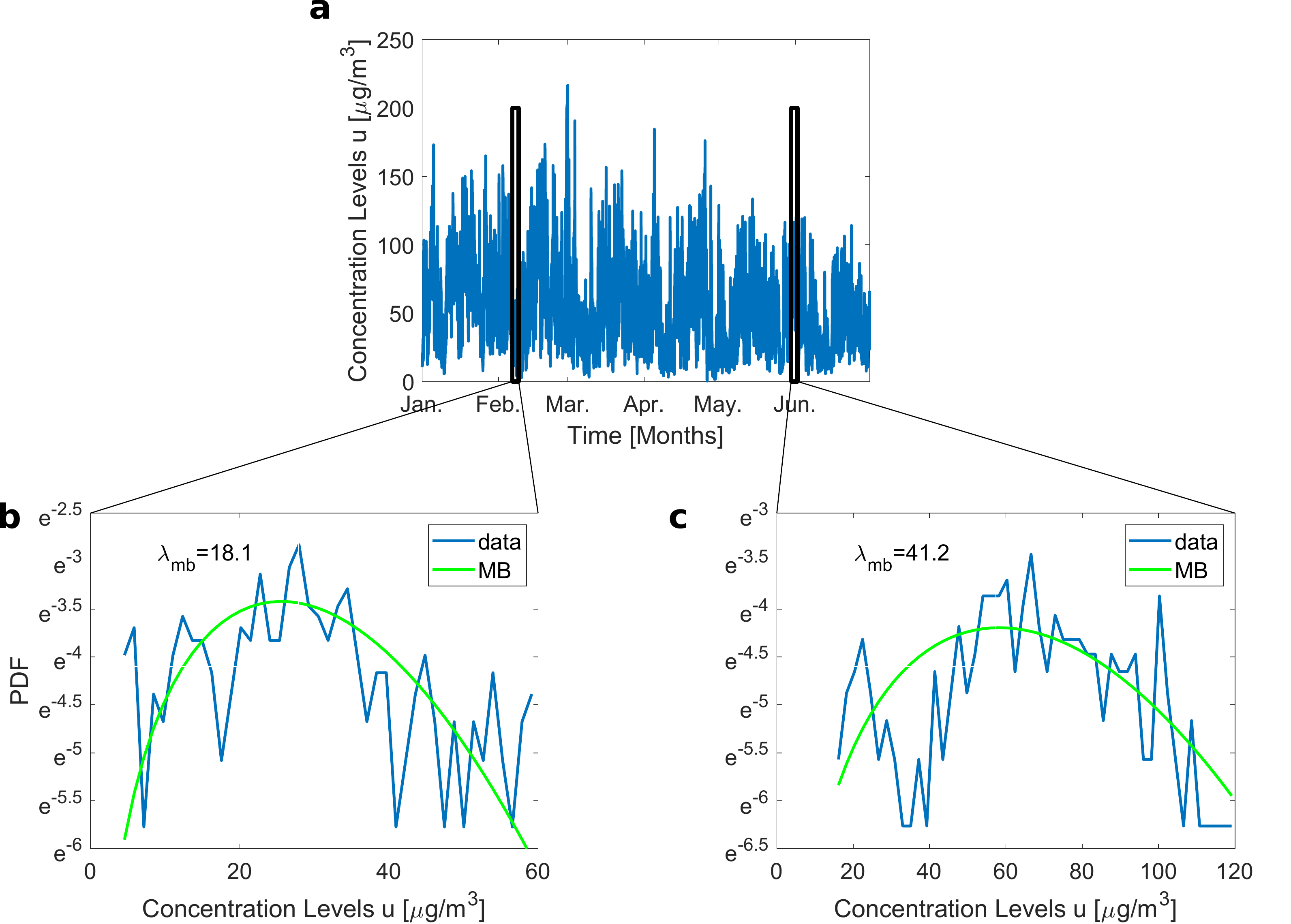}
\caption{$NO_2$ concentrations locally follow Maxwell-Boltzmann distributions. a: We plot the $NO_2$ concentration for several months in 2010, with two periods of length $\Delta t = T \approx 2.6~\text{days}$ highlighted, whose distributions are explored in panels b and c.  The histograms of panels b and c are fitted with Maxwell-Boltzmann distributions and we note their respective $\lambda_{mb}$-values, which are strongly varying in time.}
\label{FigNew}
\end{figure*}

\begin{figure}[h!]
\includegraphics[width=0.5\textwidth]{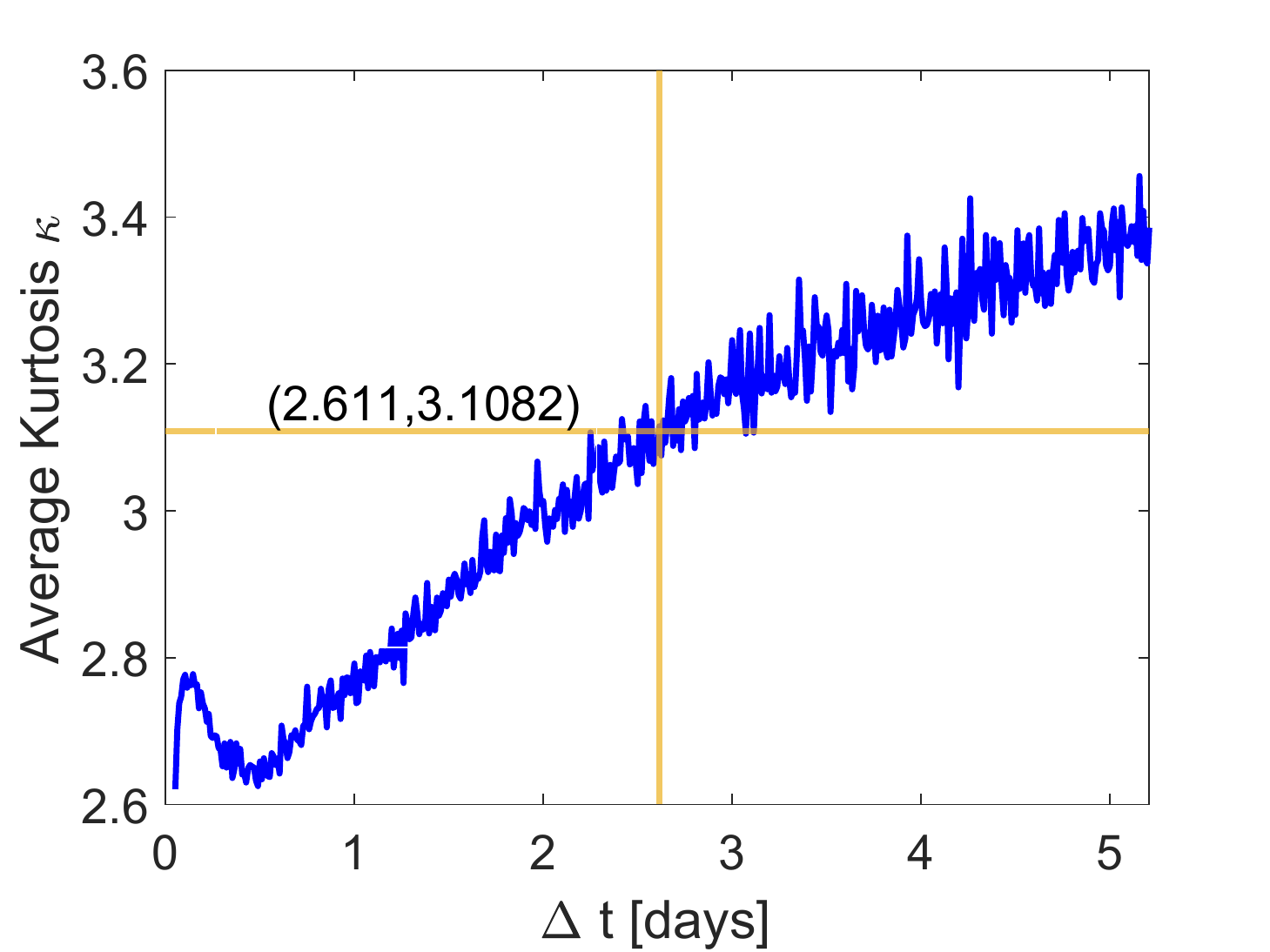}
\caption{The average kurtosis $\kappa$ is plotted as a function of the time window $\Delta t$. 
The crossing between the horizontal line at $\kappa=3.1082$ (the kurtosis of a Maxwell-Boltzmann distribution) and the $\kappa$ vs. $\Delta t$ curve gives the  value $\Delta t = T=2.61$.}
\label{Fig8}
\end{figure}
\section{Thresholds and policies} 
Policies to tackle pollution in cities often focus on thresholds and compliance with these thresholds. This then often leads to pollution just below the threshold in many regions. However, in many cases it may be more
reasonable to reduce the overall exposure to pollutants \cite{fuller2019Science}.
Weather is responsible for many fast variations of the air pollution \cite{fuller2019Science}.
The focus on threshold is especially dangerous as there is strong evidence that even small concentrations of pollutants, below national standards, increase the death rate \cite{Qian2017}.
The UK government (Dept. of Health \& Social Care) calls air pollution a "health emergency" and the World Health Organization (WHO) judges the situation similarly on a global scale \cite{WHOPollution}.
Note that about 9500 premature deaths are attributed to air pollution in London alone every year \cite{walton2015understanding}. On a global scale this number rises to about 4.9 million premature deaths, making air pollution the fifth leading cause of mortality worldwide \cite{health2018state}.

So let us finally sketch how the knowledge of the probability density function (PDF) allows estimates of threshold crossings and total exposure.
Setting thresholds on pollutant concentrations is a popular tool when setting pollution policies. Simultaneously, data coverage might not be very good in all locations. Especially if data is not available for the full time period of interest but only for a few months, estimates on how frequently thresholds are violated are difficult to obtain. As soon as we are able to determine the PDF $p(u)$ for the concentration levels $u$, we can easily compute the number of days where thresholds are violated as
\begin{equation}
            N_{u>u_\text{threshold}}=365\int_{u_\text{threshold}}^\infty p(u)\text{d}u.
\end{equation}
This integral can be easily evaluated numerically, using $q$-exponential distributions for $NO$, see Eq. (8) from the main text,
and Eq.~(17) from the main text for $NO_2$, 
consistent with experimental
observations.

The number of threshold violations explicitly depends on the parameters $\lambda_q$ and $q$, which have to be determined from the data sets but can then be compared between different locations and time periods. Having these two parameters explicitly disentangles two aspects: The rate of (rare) extreme events, encoded in $q$ and the overall variability of the pollutant concentration, encoded in $\lambda_q$ (or $\lambda_{mb}$ for $NO_2$).

Complementary to thresholds, policies could instead focus on the total pollution exposure (TPE) describing the average amount citizens are exposed to. Again, this is easily computed from the PDF as
\begin{equation}
            TPE=365\int_{0}^\infty p(u)u\text{d}u.
\end{equation}

\bibliographystyle{apsrev}
\bibliography{References}

\begin{thebibliography}{30}
\expandafter\ifx\csname natexlab\endcsname\relax\def\natexlab#1{#1}\fi
\expandafter\ifx\csname bibnamefont\endcsname\relax
  \def\bibnamefont#1{#1}\fi
\expandafter\ifx\csname bibfnamefont\endcsname\relax
  \def\bibfnamefont#1{#1}\fi
\expandafter\ifx\csname citenamefont\endcsname\relax
  \def\citenamefont#1{#1}\fi
\expandafter\ifx\csname url\endcsname\relax
  \def\url#1{\texttt{#1}}\fi
\expandafter\ifx\csname urlprefix\endcsname\relax\def\urlprefix{URL }\fi
\providecommand{\bibinfo}[2]{#2}
\providecommand{\eprint}[2][]{\url{#2}}

\bibitem[{\citenamefont{Beck and Cohen}(2003)}]{beck-cohen}
\bibinfo{author}{\bibfnamefont{C.}~\bibnamefont{Beck}} \bibnamefont{and}
  \bibinfo{author}{\bibfnamefont{E.~G.~D.} \bibnamefont{Cohen}},
  \bibinfo{journal}{Physica A: Statistical mechanics and its applications}
  \textbf{\bibinfo{volume}{322}}, \bibinfo{pages}{267} (\bibinfo{year}{2003}).

\bibitem[{\citenamefont{Beck et~al.}(2005)\citenamefont{Beck, Cohen, and
  Swinney}}]{BCS}
\bibinfo{author}{\bibfnamefont{C.}~\bibnamefont{Beck}},
  \bibinfo{author}{\bibfnamefont{E.~G.~D.} \bibnamefont{Cohen}},
  \bibnamefont{and} \bibinfo{author}{\bibfnamefont{H.~L.}
  \bibnamefont{Swinney}}, \bibinfo{journal}{Physical Review E}
  \textbf{\bibinfo{volume}{72}}, \bibinfo{pages}{056133}
  (\bibinfo{year}{2005}).

\bibitem[{\citenamefont{Beck}(2007)}]{beck2007}
\bibinfo{author}{\bibfnamefont{C.}~\bibnamefont{Beck}},
  \bibinfo{journal}{Physical Review Letters} \textbf{\bibinfo{volume}{98}},
  \bibinfo{pages}{064502} (\bibinfo{year}{2007}).

\bibitem[{\citenamefont{Daniels et~al.}(2004)\citenamefont{Daniels, Beck, and
  Bodenschatz}}]{bodenschatz}
\bibinfo{author}{\bibfnamefont{K.~E.} \bibnamefont{Daniels}},
  \bibinfo{author}{\bibfnamefont{C.}~\bibnamefont{Beck}}, \bibnamefont{and}
  \bibinfo{author}{\bibfnamefont{E.}~\bibnamefont{Bodenschatz}},
  \bibinfo{journal}{Physica D: Nonlinear Phenomena}
  \textbf{\bibinfo{volume}{193}}, \bibinfo{pages}{208} (\bibinfo{year}{2004}).

\bibitem[{\citenamefont{Rizzo and Rapisarda}(2004)}]{rapisarda}
\bibinfo{author}{\bibfnamefont{S.}~\bibnamefont{Rizzo}} \bibnamefont{and}
  \bibinfo{author}{\bibfnamefont{A.}~\bibnamefont{Rapisarda}}, in
  \emph{\bibinfo{booktitle}{AIP Conference Proceedings}}
  (\bibinfo{organization}{AIP}, \bibinfo{year}{2004}), vol.
  \bibinfo{volume}{742}, pp. \bibinfo{pages}{176--181}.

\bibitem[{\citenamefont{Weber et~al.}(2018)\citenamefont{Weber, Reyers, Beck,
  Timme, Pinto, Witthaut, and Sch{\"a}fer}}]{weber2018wind}
\bibinfo{author}{\bibfnamefont{J.}~\bibnamefont{Weber}},
  \bibinfo{author}{\bibfnamefont{M.}~\bibnamefont{Reyers}},
  \bibinfo{author}{\bibfnamefont{C.}~\bibnamefont{Beck}},
  \bibinfo{author}{\bibfnamefont{M.}~\bibnamefont{Timme}},
  \bibinfo{author}{\bibfnamefont{J.~G.} \bibnamefont{Pinto}},
  \bibinfo{author}{\bibfnamefont{D.}~\bibnamefont{Witthaut}}, \bibnamefont{and}
  \bibinfo{author}{\bibfnamefont{B.}~\bibnamefont{Sch{\"a}fer}},
  \bibinfo{journal}{arXiv preprint arXiv:1810.06391}  (\bibinfo{year}{2018}).

\bibitem[{\citenamefont{Jizba and Kleinert}(2008)}]{jizba}
\bibinfo{author}{\bibfnamefont{P.}~\bibnamefont{Jizba}} \bibnamefont{and}
  \bibinfo{author}{\bibfnamefont{H.}~\bibnamefont{Kleinert}},
  \bibinfo{journal}{Physical Review E} \textbf{\bibinfo{volume}{78}},
  \bibinfo{pages}{031122} (\bibinfo{year}{2008}).

\bibitem[{\citenamefont{Chechkin et~al.}(2017)\citenamefont{Chechkin, Seno,
  Metzler, and Sokolov}}]{metzler}
\bibinfo{author}{\bibfnamefont{A.~V.} \bibnamefont{Chechkin}},
  \bibinfo{author}{\bibfnamefont{F.}~\bibnamefont{Seno}},
  \bibinfo{author}{\bibfnamefont{R.}~\bibnamefont{Metzler}}, \bibnamefont{and}
  \bibinfo{author}{\bibfnamefont{I.~M.} \bibnamefont{Sokolov}},
  \bibinfo{journal}{Physical Review X} \textbf{\bibinfo{volume}{7}},
  \bibinfo{pages}{021002} (\bibinfo{year}{2017}).

\bibitem[{\citenamefont{Sch{\"a}fer et~al.}(2018)\citenamefont{Sch{\"a}fer,
  Beck, Aihara, Witthaut, and Timme}}]{nature-energy}
\bibinfo{author}{\bibfnamefont{B.}~\bibnamefont{Sch{\"a}fer}},
  \bibinfo{author}{\bibfnamefont{C.}~\bibnamefont{Beck}},
  \bibinfo{author}{\bibfnamefont{K.}~\bibnamefont{Aihara}},
  \bibinfo{author}{\bibfnamefont{D.}~\bibnamefont{Witthaut}}, \bibnamefont{and}
  \bibinfo{author}{\bibfnamefont{M.}~\bibnamefont{Timme}},
  \bibinfo{journal}{Nature Energy} \textbf{\bibinfo{volume}{3}},
  \bibinfo{pages}{119} (\bibinfo{year}{2018}).

\bibitem[{\citenamefont{Yalcin et~al.}(2016)\citenamefont{Yalcin, Rabassa, and
  Beck}}]{yalcin}
\bibinfo{author}{\bibfnamefont{G.~C.} \bibnamefont{Yalcin}},
  \bibinfo{author}{\bibfnamefont{P.}~\bibnamefont{Rabassa}}, \bibnamefont{and}
  \bibinfo{author}{\bibfnamefont{C.}~\bibnamefont{Beck}},
  \bibinfo{journal}{Journal of Physics A: Mathematical and Theoretical}
  \textbf{\bibinfo{volume}{49}}, \bibinfo{pages}{154001}
  (\bibinfo{year}{2016}).

\bibitem[{\citenamefont{De~Michele and Avanzi}(2018)}]{demichel}
\bibinfo{author}{\bibfnamefont{C.}~\bibnamefont{De~Michele}} \bibnamefont{and}
  \bibinfo{author}{\bibfnamefont{F.}~\bibnamefont{Avanzi}},
  \bibinfo{journal}{Scientific Reports} \textbf{\bibinfo{volume}{8}},
  \bibinfo{pages}{14204} (\bibinfo{year}{2018}).

\bibitem[{\citenamefont{Amster et~al.}(2014)\citenamefont{Amster, Haim, Dubnov,
  and Broday}}]{Amster}
\bibinfo{author}{\bibfnamefont{E.~D.} \bibnamefont{Amster}},
  \bibinfo{author}{\bibfnamefont{M.}~\bibnamefont{Haim}},
  \bibinfo{author}{\bibfnamefont{J.}~\bibnamefont{Dubnov}}, \bibnamefont{and}
  \bibinfo{author}{\bibfnamefont{D.~M.} \bibnamefont{Broday}},
  \bibinfo{journal}{Environmental Pollution} \textbf{\bibinfo{volume}{186}},
  \bibinfo{pages}{20} (\bibinfo{year}{2014}).

\bibitem[{\citenamefont{Hamra et~al.}(2015)\citenamefont{Hamra, Laden, Cohen,
  Raaschou-Nielsen, Brauer, and Loomis}}]{Hamra}
\bibinfo{author}{\bibfnamefont{G.~B.} \bibnamefont{Hamra}},
  \bibinfo{author}{\bibfnamefont{F.}~\bibnamefont{Laden}},
  \bibinfo{author}{\bibfnamefont{A.~J.} \bibnamefont{Cohen}},
  \bibinfo{author}{\bibfnamefont{O.}~\bibnamefont{Raaschou-Nielsen}},
  \bibinfo{author}{\bibfnamefont{M.}~\bibnamefont{Brauer}}, \bibnamefont{and}
  \bibinfo{author}{\bibfnamefont{D.}~\bibnamefont{Loomis}},
  \bibinfo{journal}{Environmental Health Perspectives}
  \textbf{\bibinfo{volume}{123}}, \bibinfo{pages}{1107} (\bibinfo{year}{2015}).

\bibitem[{\citenamefont{Shah et~al.}(2015)\citenamefont{Shah, Lee, McAllister,
  Hunter, Nair, Whiteley, Langrish, Newby, and Mills}}]{Shah}
\bibinfo{author}{\bibfnamefont{A.~S.} \bibnamefont{Shah}},
  \bibinfo{author}{\bibfnamefont{K.~K.} \bibnamefont{Lee}},
  \bibinfo{author}{\bibfnamefont{D.~A.} \bibnamefont{McAllister}},
  \bibinfo{author}{\bibfnamefont{A.}~\bibnamefont{Hunter}},
  \bibinfo{author}{\bibfnamefont{H.}~\bibnamefont{Nair}},
  \bibinfo{author}{\bibfnamefont{W.}~\bibnamefont{Whiteley}},
  \bibinfo{author}{\bibfnamefont{J.~P.} \bibnamefont{Langrish}},
  \bibinfo{author}{\bibfnamefont{D.~E.} \bibnamefont{Newby}}, \bibnamefont{and}
  \bibinfo{author}{\bibfnamefont{N.~L.} \bibnamefont{Mills}},
  \bibinfo{journal}{BMJ} \textbf{\bibinfo{volume}{350}}, \bibinfo{pages}{h1295}
  (\bibinfo{year}{2015}).

\bibitem[{\citenamefont{{United States Environmental Protection
  Agency}}(2016)}]{USEPA}
\bibinfo{author}{\bibnamefont{{United States Environmental Protection
  Agency}}}, \emph{\bibinfo{title}{{Nitrogen Dioxide ($NO_{2}$) Pollution}}},
  \bibinfo{howpublished}{\url{http://www.epa.gov/no2-pollution/basic-information-about-no2\$\#\$What\$\%\$20is\$\%\$20NO2}}
  (\bibinfo{year}{2016}).

\bibitem[{\citenamefont{Domingos et~al.}(2015)\citenamefont{Domingos, Prado,
  Wong, Gehring, and Feijo}}]{Dom}
\bibinfo{author}{\bibfnamefont{P.}~\bibnamefont{Domingos}},
  \bibinfo{author}{\bibfnamefont{A.~M.} \bibnamefont{Prado}},
  \bibinfo{author}{\bibfnamefont{A.}~\bibnamefont{Wong}},
  \bibinfo{author}{\bibfnamefont{C.}~\bibnamefont{Gehring}}, \bibnamefont{and}
  \bibinfo{author}{\bibfnamefont{J.~A.} \bibnamefont{Feijo}},
  \bibinfo{journal}{Molecular Plant} \textbf{\bibinfo{volume}{8}},
  \bibinfo{pages}{506} (\bibinfo{year}{2015}).

\bibitem[{\citenamefont{Astier et~al.}(2017)\citenamefont{Astier, Gross, and
  Durner}}]{Astier}
\bibinfo{author}{\bibfnamefont{J.}~\bibnamefont{Astier}},
  \bibinfo{author}{\bibfnamefont{I.}~\bibnamefont{Gross}}, \bibnamefont{and}
  \bibinfo{author}{\bibfnamefont{J.}~\bibnamefont{Durner}},
  \bibinfo{journal}{Journal of Experimental Botany}
  \textbf{\bibinfo{volume}{69}}, \bibinfo{pages}{3401} (\bibinfo{year}{2017}).

\bibitem[{\citenamefont{Musameh et~al.}(2018)\citenamefont{Musameh, Dunn,
  Uddin, Sutherland, and Rapson}}]{Must}
\bibinfo{author}{\bibfnamefont{M.~M.} \bibnamefont{Musameh}},
  \bibinfo{author}{\bibfnamefont{C.~J.} \bibnamefont{Dunn}},
  \bibinfo{author}{\bibfnamefont{M.~H.} \bibnamefont{Uddin}},
  \bibinfo{author}{\bibfnamefont{T.~D.} \bibnamefont{Sutherland}},
  \bibnamefont{and} \bibinfo{author}{\bibfnamefont{T.~D.}
  \bibnamefont{Rapson}}, \bibinfo{journal}{Biosensors and Bioelectronics}
  \textbf{\bibinfo{volume}{103}}, \bibinfo{pages}{26} (\bibinfo{year}{2018}).

\bibitem[{\citenamefont{Barsan}(2007)}]{DHHS}
\bibinfo{author}{\bibfnamefont{M.~E.} \bibnamefont{Barsan}}
  (\bibinfo{year}{2007}).

\bibitem[{\citenamefont{{Environmental Research Group (ERG) at King's College
  London}}()}]{laqn}
\bibinfo{author}{\bibnamefont{{Environmental Research Group (ERG) at King's
  College London}}}, \emph{\bibinfo{title}{London air}},
  \urlprefix\url{https://www.londonair.org.uk}.

\bibitem[{\citenamefont{Tsallis}(2009)}]{tsallis}
\bibinfo{author}{\bibfnamefont{C.}~\bibnamefont{Tsallis}},
  \emph{\bibinfo{title}{Introduction to nonextensive statistical mechanics:
  approaching a complex world}} (\bibinfo{publisher}{Springer},
  \bibinfo{year}{2009}).

\bibitem[{\citenamefont{Hanel et~al.}(2011)\citenamefont{Hanel, Thurner, and
  Gell-Mann}}]{hanel2011}
\bibinfo{author}{\bibfnamefont{R.}~\bibnamefont{Hanel}},
  \bibinfo{author}{\bibfnamefont{S.}~\bibnamefont{Thurner}}, \bibnamefont{and}
  \bibinfo{author}{\bibfnamefont{M.}~\bibnamefont{Gell-Mann}},
  \bibinfo{journal}{Proceedings of the National Academy of Sciences}
  \textbf{\bibinfo{volume}{108}}, \bibinfo{pages}{6390} (\bibinfo{year}{2011}).

\bibitem[{\citenamefont{Jizba and Korbel}(2019)}]{jizba2019}
\bibinfo{author}{\bibfnamefont{P.}~\bibnamefont{Jizba}} \bibnamefont{and}
  \bibinfo{author}{\bibfnamefont{J.}~\bibnamefont{Korbel}},
  \bibinfo{journal}{Physical Review Letters} \textbf{\bibinfo{volume}{122}},
  \bibinfo{pages}{120601} (\bibinfo{year}{2019}).

\bibitem[{\citenamefont{Beck}(2001)}]{beck2001}
\bibinfo{author}{\bibfnamefont{C.}~\bibnamefont{Beck}},
  \bibinfo{journal}{Physical Review Letters} \textbf{\bibinfo{volume}{87}},
  \bibinfo{pages}{180601} (\bibinfo{year}{2001}).

\bibitem[{\citenamefont{Chen and Beck}(2008)}]{chen2008superstatistical}
\bibinfo{author}{\bibfnamefont{L.~L.} \bibnamefont{Chen}} \bibnamefont{and}
  \bibinfo{author}{\bibfnamefont{C.}~\bibnamefont{Beck}},
  \bibinfo{journal}{Physica A: Statistical Mechanics and its Applications}
  \textbf{\bibinfo{volume}{387}}, \bibinfo{pages}{3162} (\bibinfo{year}{2008}).

\bibitem[{\citenamefont{Fuller and Font}(2019)}]{fuller2019Science}
\bibinfo{author}{\bibfnamefont{G.~W.} \bibnamefont{Fuller}} \bibnamefont{and}
  \bibinfo{author}{\bibfnamefont{A.}~\bibnamefont{Font}},
  \bibinfo{journal}{Science} \textbf{\bibinfo{volume}{365}},
  \bibinfo{pages}{322} (\bibinfo{year}{2019}).

\bibitem[{\citenamefont{Di et~al.}(2017)\citenamefont{Di, Wang, Zanobetti,
  Wang, Koutrakis, Choirat, Dominici, and Schwartz}}]{Qian2017}
\bibinfo{author}{\bibfnamefont{Q.}~\bibnamefont{Di}},
  \bibinfo{author}{\bibfnamefont{Y.}~\bibnamefont{Wang}},
  \bibinfo{author}{\bibfnamefont{A.}~\bibnamefont{Zanobetti}},
  \bibinfo{author}{\bibfnamefont{Y.}~\bibnamefont{Wang}},
  \bibinfo{author}{\bibfnamefont{P.}~\bibnamefont{Koutrakis}},
  \bibinfo{author}{\bibfnamefont{C.}~\bibnamefont{Choirat}},
  \bibinfo{author}{\bibfnamefont{F.}~\bibnamefont{Dominici}}, \bibnamefont{and}
  \bibinfo{author}{\bibfnamefont{J.~D.} \bibnamefont{Schwartz}},
  \bibinfo{journal}{New England Journal of Medicine}
  \textbf{\bibinfo{volume}{376}}, \bibinfo{pages}{2513} (\bibinfo{year}{2017}).

\bibitem[{\citenamefont{{World Health Organization
  (WHO)}}(2019)}]{WHOPollution}
\bibinfo{author}{\bibnamefont{{World Health Organization (WHO)}}},
  \emph{\bibinfo{title}{How air pollution is destroying our health}}
  (\bibinfo{year}{2019}),
  \urlprefix\url{https://www.who.int/air-pollution/news-and-events/how-air-pollution-is-destroying-our-health}.

\bibitem[{\citenamefont{Walton et~al.}(2015)\citenamefont{Walton, Dajnak,
  Beevers, Williams, Watkiss, and Hunt}}]{walton2015understanding}
\bibinfo{author}{\bibfnamefont{H.}~\bibnamefont{Walton}},
  \bibinfo{author}{\bibfnamefont{D.}~\bibnamefont{Dajnak}},
  \bibinfo{author}{\bibfnamefont{S.}~\bibnamefont{Beevers}},
  \bibinfo{author}{\bibfnamefont{M.}~\bibnamefont{Williams}},
  \bibinfo{author}{\bibfnamefont{P.}~\bibnamefont{Watkiss}}, \bibnamefont{and}
  \bibinfo{author}{\bibfnamefont{A.}~\bibnamefont{Hunt}},
  \bibinfo{journal}{London: Kings College London, Transport for London and the
  Greater London Authority}  (\bibinfo{year}{2015}).

\bibitem[{\citenamefont{{Health Effects Institute}}(2018)}]{health2018state}
\bibinfo{author}{\bibnamefont{{Health Effects Institute}}},
  \bibinfo{journal}{Special Report}  (\bibinfo{year}{2018}).

\end{thebibliography}


\end{document}